# Study of $B_s^0 \to \ell^+\ell^-\gamma$ decays in the non–universal $Z'$ model

**D. Banerjee[+], M. Kumar, P. Nayek and S. Sahoo[*]**

Department of Physics, National Institute of Technology,
Durgapur-713209, West Bengal, India
[+]E-mail: rumidebika@gmail.com, [*]E-mail: sukadevsahoo@yahoo.com

**Abstract**

We investigate $B_s^0 \to \ell^+\ell^-\gamma$ decays in a non-universal $Z'$ model derived from extension of the standard model (SM). Considering the $Z'$-mediated flavor changing neutral current (FCNC) effects we calculate the branching ratio and forward-backward asymmetry ($A_{FB}$) for $B_s^0 \to \ell^+\ell^-\gamma$ decay processes. We compare the obtained results with predictions of the SM and discuss the sensitivity of the observables to $Z'$ boson coupling parameters. We find the branching ratios are enhanced by one order from SM predictions in $Z'$ model scenario. We also observe the variation of forward-backward asymmetry with the $Z'$ boson coupling parameters portrait discrimination between NP effects and SM results.

## 1. Introduction

The rare $B$-decays [1-18] induced by flavour-changing neutral current (FCNC) are forbidden at tree level and can only take place through loop level processes within the standard model (SM). Hence, the rare $B$-decays are one of the key areas that accommodate the possibilities of physics beyond the SM. These decays are very relatively clean and very sensitive to the flavor sector of electroweak interaction. The $B_s^0 \to \ell^+\ell^-\gamma$ decays [4-18] are one of the rare $B$-decays induced by FCNC processes involving $b \to s$ quark transition. Due to the presence of an additional photon with the lepton pair in the final state these decays are free from helicity suppression. These rare decays are attributed to the fact that they occur at loop level in the SM and are described by penguin and box diagrams leading to small branching ratios. $B_s^0 \to \ell^+\ell^-\gamma$ decays can be obtained from $B_s^0 \to \ell^+\ell^-$ transition by attaching a photon to any charged internal and/or external fermion lines in it. The contributions from the attachment of photon to any charged internal line are strongly suppressed by a factor $m_b^2/m_W^2$ in Wilson coefficients and thus can be safely neglected [4-13]. When the photon is emitted from initial quarks, the structure dependent (SD) part contributes strongly for the amplitude of $B_s^0 \to \ell^+\ell^-\gamma$ decays as they are free from the helicity suppression of these radiative decay. But when the photon is emitted from external charged leptons, the internal Bremsstrahlung (IB) part gives small contribution to the total amplitude since it is proportional to lepton mass. We consider both SD and IB part into account to calculate the total amplitude for $B_s^0 \to \ell^+\ell^-\gamma$ decay modes.

The results from LHCb collaboration for $B$-system [19, 20] based on the 3fb$^{-1}$ of data show deviation from the SM predictions. The recent measurement from Belle collaboration [21] also supports the LHCb results. Further, observation of 3.7σ deviation in the angular observable $P_5'$ [20, 22-24] of $B \to K^* \mu^+ \mu^-$ mode in $q^2 \in [4.30, 8.63]\ GeV^2$ bin, observation of 2.6σ deviation in $R_K = BR(B^+ \to K^+\mu^+\mu^-)/BR(B^+ \to K^+e^+e^-)$ in the $q^2 \in [1, 6]$ GeV$^2$ bin from SM prediction [25], for $R_{K^*} = \mathcal{B}(B \to K^*\mu^+\mu^-)/\mathcal{B}(B \to K^*e^+e^-)$ [26] the discrepancy is 2.1σ -2.3σ for low-q$^2$ region and 2.4σ-2.5σ for central-q$^2$ region from the SM prediction, discrepancy in the differential branching fraction of the $B \to K^* \ell^+ \ell^-$ processes [27-28] by the LHCb experiment, the observation of 3.2σ deviation in the decay rate of the $B_s \to \phi \mu^+ \mu^-$ [29] process and many more experimental results [30-35] illustrate the anomalies with the SM predictions. Though these deviations are not statistically satisfactory to prove the presence of NP effects but these data have intimated several anomalies in $B$ decays induced by FCNC processes $b \to s\,\ell^+\ell^-$ which demands new physics models as well as model independent way to explain the source of these anomalies vigorously. Again as $b \to s$ transitions occur via loop processes, it is possible to introduce new particles to execute these transitions. Several studies are done on $b \to s$ transitions with models beyond the SM [4-6, 36-44]. So, it is necessary to search these decays by direct or indirect approach in recent time to visualize the physics beyond the SM horizon. Dettori *et al.* [14] have conferred a novel strategy to search for the $B_S \to \mu^+\mu^-\gamma$ in the event sample selected for $B_S \to \mu^+\mu^-$ searches. New physics effects in $B_S^0 \to \ell^+\ell^-\gamma$ decays can be visualized in two ways: one is through new contributions to the existing Wilson coefficients of SM or through modifying the effective Hamiltonian in the SM.

The $B_S^0 \to \ell^+\ell^-\gamma$ decays are sensitive to the existence of new physics (NP) beyond the SM. Non-universal $Z'$ model is one of the possible extensions of the SM [45-49] which contains extra $U(1)'$ gauge symmetry. The $Z'$ boson is not found experimentally so far, hence the exact mass of the $Z'$ boson is not known. However, there are stringent limits on the mass of $Z'$ boson and the $Z - Z'$ mixing angle $\theta$ from the non-observation of direct production at the CDF [50-53] and indirect constraints from the precision data (weak neutral current processes and LEP II) [54-57]. Mixing between ordinary and exotic left-handed quarks induces Z-mediated FCNCs. The right-handed quarks $d_R, s_R$ and $b_R$ have different $U(1)'$ quantum numbers than exotic $q_R$. The mixing among these right-handed quarks and exotic quarks will induce $Z'$-mediated FCNCs [58-64] among the ordinary down quark types. Tree level FCNC interactions can also be induced by an additional $Z'$ boson due to the non-diagonal chiral coupling matrix. With FCNCs, both Z and $Z'$ boson contributes at tree level and it will interfere with the SM contributions [61-66]. Many studies of non-universal $Z'$ model have been performed by various authors in order to fix the observed anomalies of rare B meson decays and in $B_q - \overline{B_q}$ mixing [41-44]. In [67] Altmannshofer *et al.* have explained that the long-distance deviation in the muon g-2 anomaly can be justified in NP scenario with heavy $Z'$ gauge boson with $m_{Z'} \geq 400\ MeV$. The authors have also pointed out in [68] that the heavy $Z'$ boson gives additional contribution towards the tests of physics beyond the SM in the process of neutrino trident production. In the early 90s' the direct searches from different accelerators gave model dependent lower bounds around 500 GeV for $m_{Z'}$ [50-52]. The major detection technique for the $Z'$ boson is Drell-Yan production channel of dilepton resonance $pp \to Z' \to \ell^+\ell^- + X$ [69-71] at the LHC which gives constraints on $m_{Z'}$ along with $Z - Z'$ mixing angle $\theta_0$ and coupling parameter $g'$. By analysing the current LHC Drell-Yan data, Bandyopadhyay *et al.* [71] have obtained $m_{Z'} > 4.4\, TeV$ and the mixing angle $\theta_0 < 10^{-3}$ by considering the strength of the $U(1)'$ gauge coupling is same as that of the SM $SU(2)_L$. ATLAS collaboration constrained $m_{Z'} > 2.42\, TeV$ [69] for the sequential standard model (SSM) and $m_{Z'} > 4.1\, TeV$ [72] for the $E_6$ motivated $Z'$ model. Recently CMS collaboration have predicted the lower limit on $m_{Z'}$ as 4.5 TeV in SSM and 3.9 TeV in superstring-inspired model by studying the decay of $Z'$ into dilepton final state ($e^+e^-$ and $\mu^+\mu^-$) in *p-p* collision with $\sqrt{s} = 13\, TeV$. The upper bound $m_{Z'} \leq 6\ TeV$ have been predicted in a classically conformal $U(1)'$ extended standard model in [73]. In low energy experiments [74] the model dependent lower limit of $m_{Z'}$ are above 1 TeV. An extensive analysis of $B_s^0 \to \ell^+\ell^-\gamma$ decays in a family non-universal $Z'$ model in different $q^2$ regions has been done by Sirvanh [75]. He found that the measurable quantities (e.g. differential branching ratio, double lepton polarizations and forward-backward asymmetries) are quite sensitive to the $Z'$ contributions especially in low $q^2$ region and give indication for the existence of the $Z'$ boson. In this paper, we focus on studying the contribution of $Z'$ boson coupling parameters to the observables like branching ratio and forward backward asymmetry of $B_s^0 \to \ell^+\ell^-\gamma(\ell = \mu, \tau)$ decays and compare our results with the SM predictions. To investigate the $B_s^0 \to \ell^+\ell^-\gamma$ decay it is essential to parameterize the decay matrix elements in terms of invariant form factors. Theoretical estimation to obtain the associated form factors satisfying the constraints for different range of photon energy have already been carried out in different approaches by many authors which includes perturbative QCD methods [76], the light-cone QCD sum rule technique [4-7], constituent quark model [77], light front model [78], dispersion approach [79-80], large energy effective theory [81], soft collinear effective theory [82]. Authors in [83] have computed the structure dependent part of the matrix element $B_s^0 \to \ell^+\ell^-\gamma$ $(\ell = e, \mu)$ by considering the universal form of independent form factors which were computed [76] using perturbative QCD methods combined with heavy quark effective theory along with the light cone wave function of the B meson. In [84] the universality of non-perturbative QCD effects in different radiative B meson decays including $B_s^0 \to \ell^+\ell^-\gamma$ decay have been studied by an explicit one loop calculation. Long distance QCD effects were computed in [10] for $B_s^0 \to \ell^+\ell^-\gamma$ decays, which shows the presence of light vector-meson resonances related to the virtual photon emission from external quark lines of B meson decay contributes reasonably to the dilepton differential distribution. The authors of [13] re-evaluated the rare $B_s^0 \to \ell^+\ell^-\gamma$ decays in the SM using the $B_s$ wave function extracted from non-leptonic $B_s$ decays by including the contributions from magnetic penguin operator $b \to s\gamma$ with real photons which were considered to be negligible in other literature. They found these contributions are large and enhance the branching ratios of these rare decays by a factor $\approx 2$. In recent time Kozachuk *et al.* [15] have obtained $B \to \gamma$ transition form factors in relativistic dispersion approach based on the constituent quark picture and have also analysed the charm loop contribution explicitly. In [18] the $B_s^0 \to \ell^+\ell^-\gamma$ rare decay were studied in the framework of covariant confined quark model developed in [85] based on relativistic Lagrangians describing the hadrons interaction with their constituent quarks. They have estimated $B_s \to \gamma$ transition form factors in the full kinematical region along with their values at zero recoil nearly at 10% theoretical error. In particular for our study, we have undertaken the structure of the form factors for $B_s^0 \to \ell^+\ell^-\gamma$ weak transitions in the resonance region obtained from the gauge invariance requirement of the $B_s^0 \to \ell^+\ell^-\gamma$ amplitude at large photon energy [8-10, 15-17, 86-89]. We consider these form factors are most economical for our study as it satisfies the constraints on the behavior of the form factors over the entire energy range.

This paper is organised as follows. In Section 2, we present effective Hamiltonian responsible for the $b \to s\,\ell^+\ell^-$ transitions and the matrix element for the decay modes $B_s^0 \to \ell^+\ell^-\gamma$ in the SM and discuss the forward backward asymmetry associated with the final state leptons. In Section 3, we discuss the effect of the $Z'$-mediated FCNCs. We write the effective Hamiltonian for the $Z'$ part for $b \to s\ell^+\ell^-$ transitions following the modified Wilson coefficients $C_9$ and $C_{10}$. In Section 4, we obtain the branching ratios (BR) for $B_s^0 \to \ell^+\ell^-\gamma(\ell = \mu, \tau)$ processes in the $Z'$ model by using $Z'$ boson coupling parameters whose values are constrained from $B$ meson mixing and different inclusive as well as exclusive decays of $B$ meson. We discuss the forward backward asymmetry for the $B_s^0 \to \ell^+\ell^-\gamma(\ell = \mu, \tau)$ decay modes in the $Z'$ model and compare it with the SM predictions. Concluding remarks are presented in Section 5.

## 2. $B_s^0 \to \ell^+\ell^-\gamma$ Decay in the SM

The matrix element for the process $B_s^0 \to \ell^+\ell^-\gamma$ can be obtained from that of the pure leptonic $B_s^0 \to \ell^+\ell^-$ decay (based on $b \to s\ell^+\ell^-$quark level transition) by attaching a photon to any charged external fermion lines, i.e. direct emission of photon from valence quark and $\ell^+\ell^-$ lepton pair (The transition is shown in Fig. 1). The quark level process $b \to s\ell^+\ell^-$ can be described by the effective Hamiltonian [8, 86],

$$H_{eff}^{b\to s\ell^+\ell^-} = \frac{G_F\alpha_{em}}{2\sqrt{2}\pi} V_{tb}V_{ts}^* \begin{bmatrix} C_9^{eff}\left(\bar{s}\gamma_\mu P_L b\right)\cdot\left(\bar{\ell}\gamma^\mu\ell\right) + C_{10}\left(\bar{s}\gamma_\mu P_L b\right)\cdot\left(\bar{\ell}\gamma^\mu\gamma_5\ell\right) \\ -\frac{2C_7^{eff} m_b}{q^2}\left(\bar{s}\, i\, \sigma_{\mu\vartheta} q^\vartheta P_R b\right)\cdot\left(\bar{\ell}\gamma^\mu\ell\right) \end{bmatrix}, \quad (1)$$

where $G_F$ is the Fermi coupling constant, $P_{L,R} = \frac{1}{2}(1 \mp \gamma_5)$, $C_i$'s are the Wilson coefficients evaluated at the $b$ quark mass scale in next-to-leading logarithm order QCD correction [86-91]. $C_9^{eff}$ and $C_{10}$ are associated with the electroweak penguin semi-leptonic operators $O_9$ and $O_{10}$ respectively and $C_7^{eff}$ is the coefficient of the electromagnetic penguin operator [86-90, 92-109] and are given as $C_7^{eff} = -0.315$, $C_9 = 4.227, C_{10} = -4.642$. Along with the short distance effect $C_9^{eff}$also receives long distance contribution which has a perturbative part Y(s) coming from one loop matrix elements of four quark operators and a resonance part $C_9^{res}$ prescribed in Breit-Wigner form. Y(s) part and $C_9^{res}$ part have been discussed in Ref. [90, 93-94, 104] and [87-89, 105-110] in detail. Hence, after introducing all these contributions $C_9^{eff}$ can be written as,

$$C_9^{eff} = C_9 + Y(s) + C_9^{res}. \quad (2)$$

The effective Hamiltonian describing the $b \to s\gamma$ electroweak transition (emission of real photon from the magnetic penguin vertex) is given as [10],

$$H_{eff}^{b\to s\gamma} = -\frac{G_F}{\sqrt{2}} V_{tb} V_{ts}^* C_7^{eff} \frac{e}{8\pi^2} m_b \left(\bar{s}\, \sigma_{\mu\vartheta} P_R b\right) F^{\mu\vartheta} \quad (3)$$

The two basis operators in equation (1) and (3) contribute to the structure dependent (SD) part of the decay amplitude. To parametrize the SD part of $B_s^0 \to \ell^+\ell^-\gamma$ decay amplitude we need a set of invariant form factors. For our calculation, we use the $B_s \to \gamma$ transition of the corresponding basis operators defined in Ref. [8-10, 87-89] as,

$$\langle \gamma(k,\epsilon) | \bar{s}\gamma_\mu \gamma_5 b | B_s(p_B) \rangle = ie[\epsilon_\mu^*(p_B.k) - (p_B \cdot \epsilon^*) k_\mu] \frac{F_A}{m_{B_s}}, \quad (4)$$

$$\langle \gamma(k,\epsilon) | \bar{s}\gamma_\mu b | B_s(p_B) \rangle = e\epsilon^{*\alpha} \epsilon_{\mu\alpha\rho\sigma} p_B^\rho k^\sigma \frac{F_V}{m_{B_s}}, \quad (5)$$

$$\langle \gamma(k,\epsilon) | \bar{s}\sigma_{\mu\vartheta} q^\vartheta \gamma_5 b | B_s(p_B) \rangle = e[\epsilon_\mu^*(p_B.k) - (p_B \cdot \epsilon^*) k_\mu] F_{TA}, \quad (6)$$

$$\langle \gamma(k,\epsilon) | \bar{s}\sigma_{\mu\vartheta} q^\vartheta b | B_s(p_B) \rangle = ie\epsilon^{*\alpha} \epsilon_{\mu\alpha\rho\sigma} p_B^\rho k^\sigma F_{TV}, \quad (7)$$

*Fi*'s depends upon two factors: the square of emitted photon momentum $k^2$ and the square of momentum transfer $q^2 = (p_B - k)^2$. The form factors $F_A, F_V, F_{TA}$ and $F_{TV}$ induced by the vector, axial vector, tensor and pseudotensor quark currents are evaluated in relativistic dispersion approach based on the constituent quark picture. The $q^2$ dependence of $F_A, F_V, F_{TA}$ and $F_{TV}$ are given as:

$$F(E_\gamma) = \beta \frac{f_{B_s} m_{B_s}}{\Delta + E_\gamma}, \quad (8)$$

where, $E_\gamma$ is the photon energy and $f_{B_s}$ is the leptonic decay constant. This is related to the dilepton invariant mass (in the *B* meson rest frame) as:

$$E_\gamma = \frac{m_{B_s}}{2}\left(1 - \frac{q^2}{m_{B_s}^2}\right). \quad (9)$$

Table-1: The parameters for $B \to \gamma$ form factors [9].

| | $F_V$ | $F_{TV}$ | $F_A$ | $F_{TA}$ |
|---|---|---|---|---|
| $\beta$ | 0.28 | 0.30 | 0.26 | 0.33 |
| $\Delta$ | 0.04 | 0.04 | 0.30 | 0.30 |

Therefore, the matrix element for the SD part corresponding to the photon emitted from the initial quarks expressed as,

$$M_{SD} = \frac{\alpha^{3/2} G_F}{\sqrt{2}\pi} V_{tb} V_{ts}^* \begin{bmatrix} \epsilon^{*\alpha} \epsilon_{\mu\alpha\rho\sigma} p_B^\rho k^\sigma \left(A_1 \bar{\ell}\gamma^\mu \ell + A_2 \bar{\ell}\gamma^\mu \gamma_5 \ell\right) + \\ i(\epsilon_\mu^*(p_B \cdot k) - (\epsilon^* \cdot p_B) k_\mu)\left(B_1 \bar{\ell}\gamma^\mu \ell + B_2 \bar{\ell}\gamma^\mu \gamma_5 \ell\right) \end{bmatrix}$$

(10)

where, $$A_1 = 2C_7^{eff} \frac{m_b}{q^2} F_{TV} + C_9^{eff} \frac{F_V}{m_{B_s}}, \quad A_2 = C_{10} \frac{F_V}{m_{B_s}}$$

$$B_1 = -2C_7^{eff} \frac{m_b}{q^2} F_{TA} - C_9^{eff} \frac{F_A}{m_{B_s}}, \quad B_2 = -C_{10} \frac{F_A}{m_{B_s}} \quad .$$

(11)

The matrix element for the IB part (structure independent) corresponding to the photon, emitted from the external charged leptons (the IB part is proportional to lepton mass, its contribution to the total matrix element is small compared to the SD part), which follows from helicity arguments. The IB part of the amplitude is described by Fig. 2 diagrams. Here, only the operator $O_{10}$ contribute to the IB part of the decay amplitude and the corresponding matrix element is given as [8-10, 90]:

$$M_{IB} = \frac{\alpha^{3/2} G_F}{\sqrt{2\pi}} V_{tb} V_{ts}^{*} f_{B_s} m_l C_{10} \left[ \overline{\ell} \left( \frac{\notin^{*} \not{p}_B}{p_{+} \cdot k} - \frac{\not{p}_B \notin^{*}}{p_{-} \cdot k} \right) \gamma_5 \ell \right]. \tag{12}$$

Hence, the total matrix element can be obtained by adding the two parts given in equation (10) and (12) together, i.e. $M = M_{SD} + M_{IB}$ and hence squared matrix element is,

$$|M|^2 = |M_{SD}|^2 + |M_{IB}|^2 + 2Re(M_{SD} M_{IB}^{*}) \tag{13}$$

The corresponding differential decay width of the $B_s^0 \to \ell^+ \ell^- \gamma$ becomes,

$$\frac{d\Gamma}{d\hat{s}} = \frac{G_F^2 \alpha^3}{2^{10} \pi^4} |V_{tb} V_{ts}^{*}|^2 m_{B_s}^3 \boldsymbol{\Delta} \tag{14}$$

where,

$$\Delta = \frac{4}{3} m_{B_s}^2 (1 - \hat{s})^2 v^l \left( (\hat{s} + 2r_l)(|A_1|^2 + |B_1|^2) + (\hat{s} - 4r_l)(|A_2|^2 + |B_2|^2) \right) - 64 \frac{f_{B_s}^2}{m_{B_s}^2} \frac{r_l}{1-\hat{s}} C_{10}^2 \left( (4r_l - \hat{s}^2 - 1) ln \frac{1-v_l}{1-v_l} + 2\hat{s}\, v_l \right) - 32 r_l (1 - \hat{s})^2 f_{B_s} Re(C_{10} A_1^{*}) \tag{15}$$

with $s = q^2$(momentum transferred square) and $4m_l^2 \leq s \leq m_{B_s}^2$, $\hat{s} = s/m_{B_s}^2$, $r_l = m_l^2/m_{B_s}^2$, and $v_l = \sqrt{1 - 4m_l^2/q^2}$.

The forward-backward (FB) asymmetry associated with the final state lepton is defined as,

$$A_{FB}(\hat{s}) = \frac{\int_0^1 \frac{d^2\Gamma}{d\hat{s}\, d\cos\theta} d\cos\theta - \int_{-1}^0 \frac{d^2\Gamma}{d\hat{s}\, d\cos\theta} d\cos\theta}{\int_0^1 \frac{d^2\Gamma}{d\hat{s}\, d\cos\theta} d\cos\theta + \int_{-1}^0 \frac{d^2\Gamma}{d\hat{s}\, d\cos\theta} d\cos\theta}\ , \tag{16}$$

where $\theta$ is the angle between momentums of $B_s$ meson and lepton in the dileptonic CM frame. Hence using the above definition, we can get the forward-backward asymmetry,

$$A_{FB} = \frac{1}{\Delta} \left\{ \begin{array}{c} 2m_{B_s}^2 \hat{s} (1 - \hat{s})^3 v_l^2 Re(A_1^{*} B_2 + B_1^{*} A_2) \\ +32 f_{B_s} r_l (1 - \hat{s})^2 \, ln\left(\frac{4r_l}{\hat{s}}\right) Re(C_{10} B_2^{*}) \end{array} \right\}. \tag{17}$$

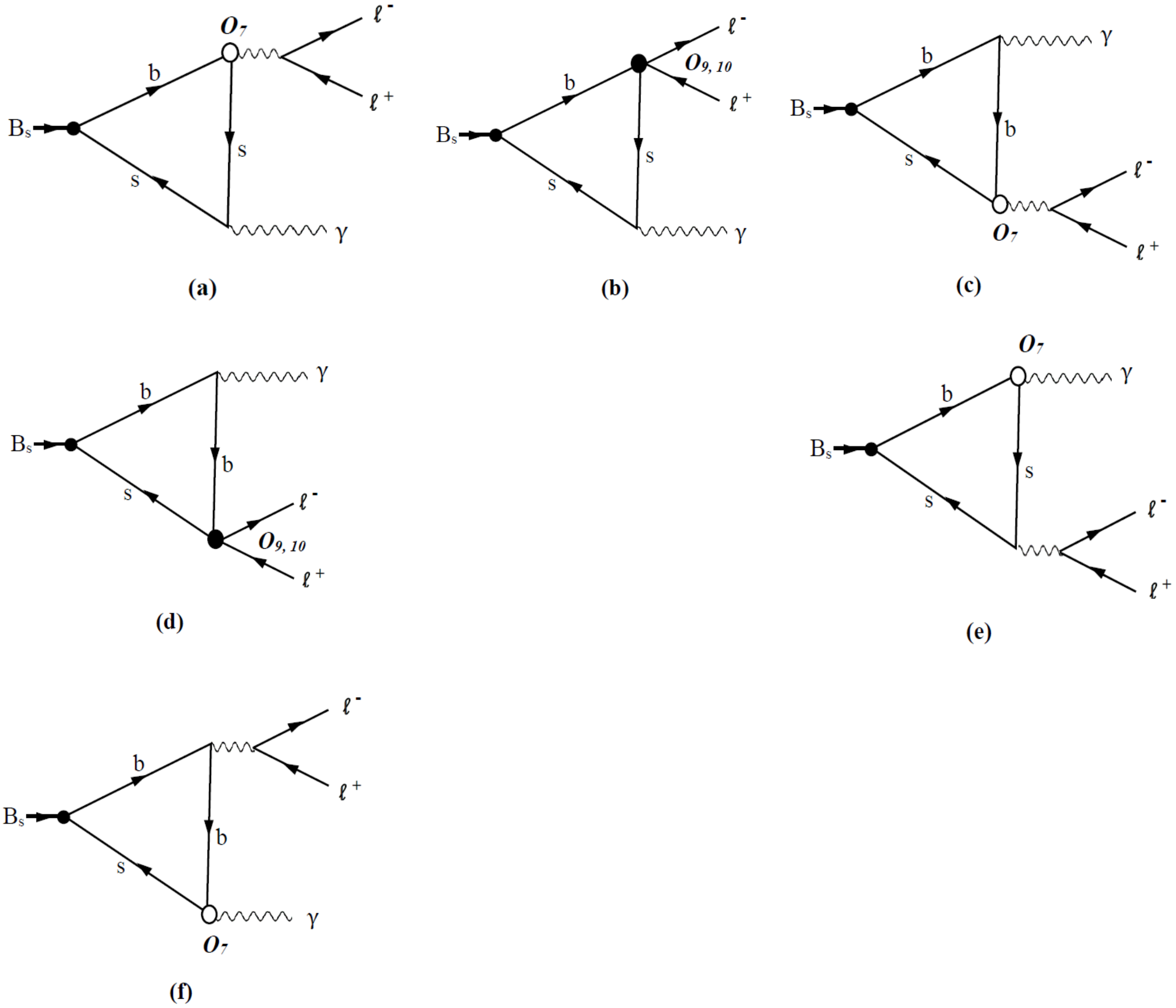

Fig. 1: Diagrams contributing to structure dependendent part of $B_s^0 \to \ell^+\ell^-\gamma$ decay. Solid circles denote the $b \to s\ell^+\ell^-$ operators $O_9$ and $O_{10}$. Hollow circles denote the $b \to s\gamma$ operator $O_7$.

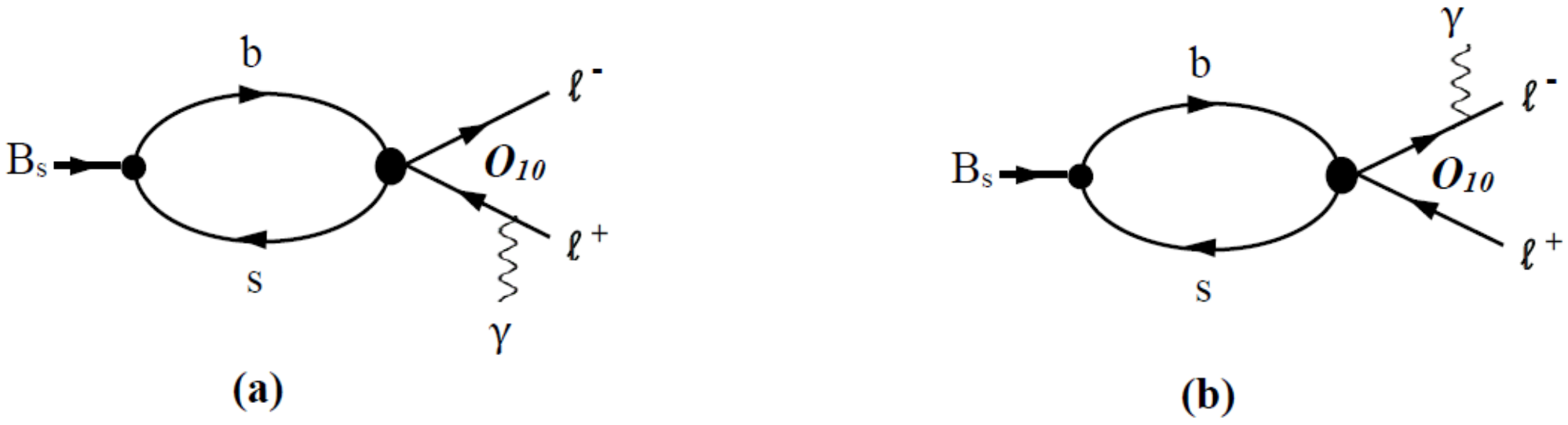

Fig. 2: Diagrams contributing to Bremsstrahlung part of $B_s^0 \to \ell^+\ell^-\gamma$ decay. Bigger solid circles denote the $b \to s\ell^+\ell^-$ operators $O_{10}$.

### 3. $B_s^0 \to \ell^+\ell^-\gamma$ Decays in the $Z'$ model

Several new physics models have been coined by many groups of scientists trying to overcome the loop holes of the SM of elementary particles and explain the source of anomalies recorded in the past few years especially for $b \to s\mu^+\mu^-$ transition. Since, the transitions $b \to s\ell^+\ell^-$ are allowed through loop processes within the SM, so an elementary approach to search NP effects in these decay modes is to establish tree level FCNCs for $b \to s\ell^+\ell^-$transitions. A family non-universal $Z'$ model [45-46] can be derived from the extension of SM by including an additional $U'(1)$ gauge symmetry to it. In this model, the FCNC transitions could be induced at tree level because of the non-diagonal chiral coupling matrix. Ignoring $Z - Z'$ mixing and considering the couplings of right-handed quark flavours with the $Z'$ boson are diagonal, the $Z'$ boson part of the effective Hamiltonian for $b \to s\ell^+\ell^-$ can be written as [64-65, 91, 111-115]:

$$H_{eff}^{Z\prime} = -\frac{4G_F}{\sqrt{2}} V_{tb}V_{ts}^*\left[\Lambda_{sb} C_9^{Z'} O_9 + \Lambda_{sb} C_{10}^{Z'} O_{10}\right] \tag{18}$$

where, $$\Lambda_{sb} = \frac{4\pi e^{-i\varphi_{sb}}}{\alpha V_{tb}V_{tb}^*} \tag{19}$$

$$C_9^{Z'} = \Lambda_{sb}|B_{sb}|S_{ll}, C_{10}^{Z'} = \Lambda_{sb}|B_{sb}|D_{ll} \tag{20}$$

The $S_{ll}$ and $D_{ll}$ are associated with the couplings of the $Z'$ boson with the left- and right-handed leptons respectively, $B_{sb}$ corresponds to the off-diagonal left-handed coupling of quarks with $Z'$ boson and $\varphi_{sb}$ corresponds to a new weak phase.

The most useful feature of the non-universal $Z'$ boson is that the operator basis remains same as in the SM and the only change occurs in Wilson coefficients $C_9$ and $C_{10}$. Hence, the effect of non-universal $Z'$ boson can be evaluated by replacing the SM Wilson coefficients $C_9$ and $C_{10}$ as:

$$C_9^{Total} = C_9^{eff} - \frac{4\pi e^{-i\varphi_{sb}}}{\alpha V_{tb}V_{tb}^*}|B_{sb}|S_{ll}, \tag{21}$$

$$C_{10}^{Total} = C_{10}^{SM} + \frac{4\pi e^{-i\varphi_{sb}}}{\alpha V_{tb}V_{tb}^*}|B_{sb}|D_{ll}. \tag{22}$$

In this model, the new physics contributions to branching ratio and forward-backward asymmetry for $B_s^0 \to \ell^+\ell^-\gamma(\ell = \mu, \tau)$ decays are analysed in the light of above modifications in Section 4.

## 4. Numerical Analysis and Discussion

In this section, we calculate the branching ratio and analyze the forward-backward asymmetry for $B_s^0 \to \ell^+\ell^-\gamma(\ell = \mu, \tau)$ decays in a non-universal $Z'$ model. For this we need to fix all input parameters. The parameters ( $\beta$ and $\Delta$) appearing in the form factors (equation (9)) are summarised in Table-1 [9]. Then to evaluate different observables in $Z'$model we need to fix the numerical values of the $Z'$coupling parameters $|B_{sb}|$, $\varphi_{sb}$, $S_{ll}$ and $D_{ll}$. The values are strictly constrained from $B$ meson mixing and different inclusive as well as exclusive decays of $B$ meson. The D0 and CDF collaborations have reported evidence for anomalously large CP violation in like-sign dimuon charge asymmetry in semileptonic decays of $b$ hadrons [116-117]. To explain the observed anomalies for the same-sign dimuon charge asymmetry and other mixing parameters, explicit study have been performed in [118-119] by considering the new contributions coming from family non-universal $Z'$ boson model. The $Z'$ model could not simultaneously reconcile all the data on $B_q^0 - \overline{B_q^0}$ mixing but it provides new constraints over the $Z'$ parameters. Here, we have considered two scenarios ($S_1$ and $S_2$ listed in Table-2) for numerical values of the $Z'$ coupling parameters corresponding to different fitting values for $B_s - \overline{B_s}$ mixing data from the UTfit Collaboration [120-121], from ref. [122-125] and from the available data of $\Delta m_s$ and branching ratios for $\overline{B_s} \to \mu^+\mu^-$ and $\overline{B_q} \to V_q l^+ l^-$studied in [126]. The numerical values of all other parameters are taken from Particle Data Group [127].

Table-2: The numerical values of $Z'$coupling parameters [91, 120-126].

| | $\lvert B_{sb} \rvert \times 10^{-3}$ | $\varphi_{sb}$ in degree | $S_{ll} \times 10^{-3}$ | $D_{ll} \times 10^{-3}$ |
|---|---|---|---|---|
| $S_1$ | $1.09 \pm 0.22$ | $72 \pm 7$ | $-2.8 \pm 3.9$ | $-6.7 \pm 2.6$ |
| $S_2$ | $2.20 \pm 0.15$ | $82 \pm 4$ | $-1.2 \pm 1.4$ | $-2.5 \pm 0.9$ |

Using all the input parameters discussed above we calculate the BR for $B_s^0 \to \ell^+\ell^-\gamma$ with $Z'$ parameters from $S_1$ and $S_2$ scenario simultaneously. In Fig. 3, we have shown variation of BR for $B_s^0 \to \ell^+\ell^-\gamma$ $(\ell = \mu, \tau)$ with respect to $Z'$ parameters within the allowed ranges in $S_1$ and $S_2$. The maximum BR corresponding to each decay in $S_1$ and $S_2$ are collectively given in Table-3.

Aliev *et al.* [4 -7] calculated the SM branching ratios, BR $(B_s^0 \to \mu^+\mu^-\gamma) = 1.9 \times 10^{-9}$ and BR $(B_s^0 \to \tau^+\tau^-\gamma) = 9.54 \times 10^{-9}$. Recently, Kozachuk *et al.* [15-17] have studied the rare radiative leptonic decays $B_s^0 \to \ell^+\ell^-\gamma$ induced by FCNCs in the SM by considering long distance QCD effects in the frame work of relativistic quark model and estimated BR$(B_s^0 \to \ell^+\ell^-\gamma) = (6.01 \pm 0.008) \times 10^{-9}$. They have also performed a detailed analysis of the charm loop contributions to these radiative decay modes. Dubnicka *et al.* [18] have studied $B_s^0 \to \ell^+\ell^-\gamma$ decays in the SM using covariant quark model and estimated BRs with (in brackets) and without long distance contributions as: BR $(B_s^0 \to \mu^+\mu^-\gamma) = 1.6(1.5) \times 10^{-9}$ and BR $(B_s^0 \to \tau^+\tau^-\gamma) = 12.7(12.6) \times 10^{-9}$. Comparing the SM predictions with our estimated BRs in the $Z'$ model given in Table-3 we can see the BR in both the scenario $S_1$ and $S_2$ are enhanced with respect to their SM predictions. Fig. 3 shows clear distinction between the predicted BR in SM and $Z'$ model.

Table-3: Numerical estimation of the BR of $B_s^0 \to \ell^+\ell^-\gamma$ $(\ell = \mu, \tau)$ in $Z'$model.

| $Z'$model | BR $(B_s^0 \to \mu^+\mu^-\gamma)$ | BR $(B_s^0 \to \tau^+\tau^-\gamma)$ |
|---|---|---|
| $S_1$ | $1.65 \times 10^{-8}$ | $1.86 \times 10^{-8}$ |
| $S_2$ | $1.08 \times 10^{-8}$ | $1.14 \times 10^{-8}$ |

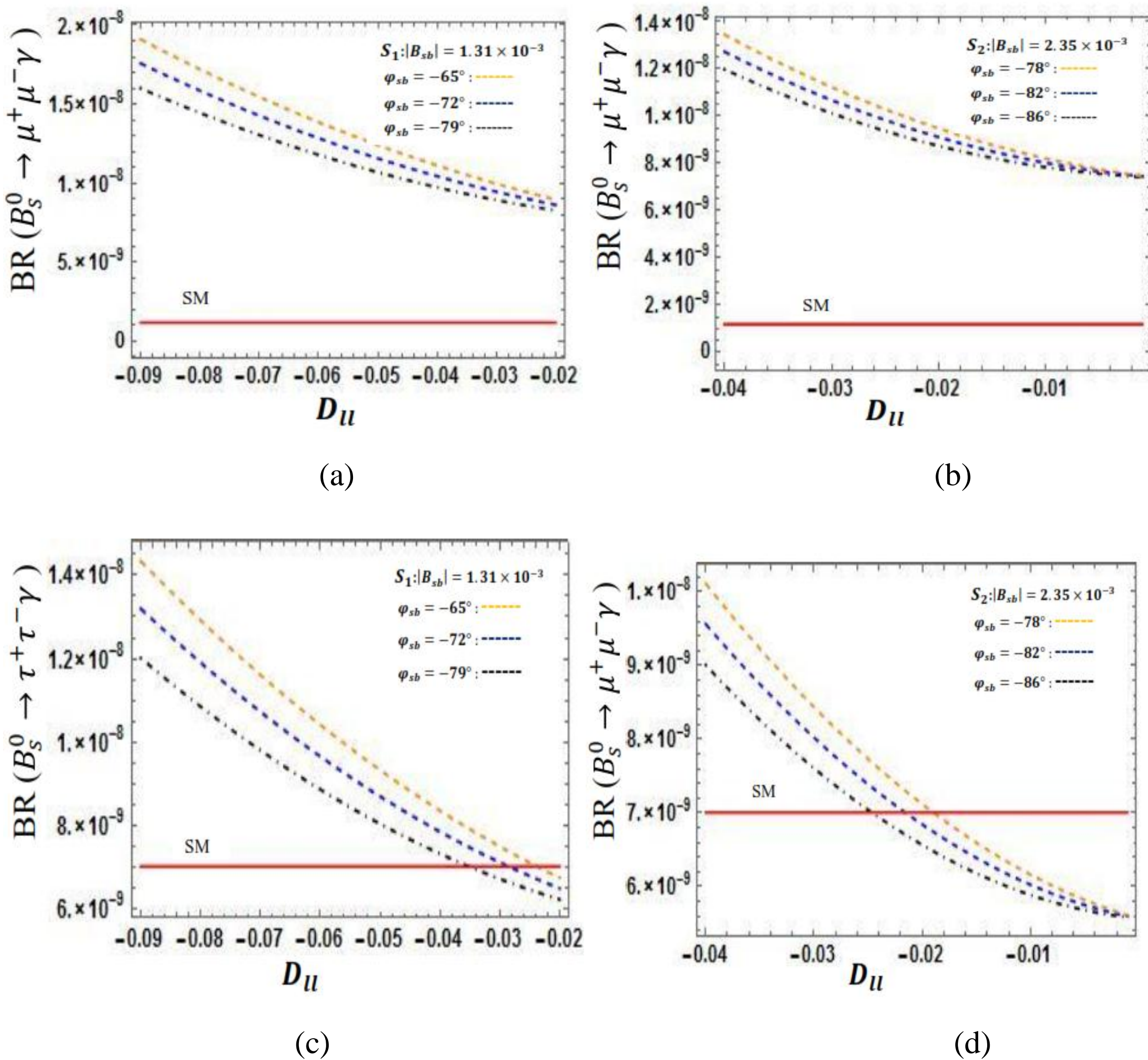

(a) (b)

(c) (d)

Fig. 3: BR ( $B_s^0 \to l^+l^-\gamma$ $(l = \mu, \tau)$ ) in $S_1$ and $S_2$ with different $\varphi_{sb}$ values.

Further, the NP contributions to the FB asymmetry are encoded in the modified Wilson coefficient. Therefore, we investigate the variation of FB asymmetry for $B_s^0 \to \ell^+\ell^-\gamma$ $(\ell = \mu, \tau)$ decays with different values of $Z'$ coupling parameter within the kinematical accessible physical range of $s$. We find that the FB asymmetry can be a good discriminant of NP. The plots shown in Fig. 4, Fig. 5, Fig. 6 and Fig. 7 certainly distinguish between NP contributions from that of SM predictions.

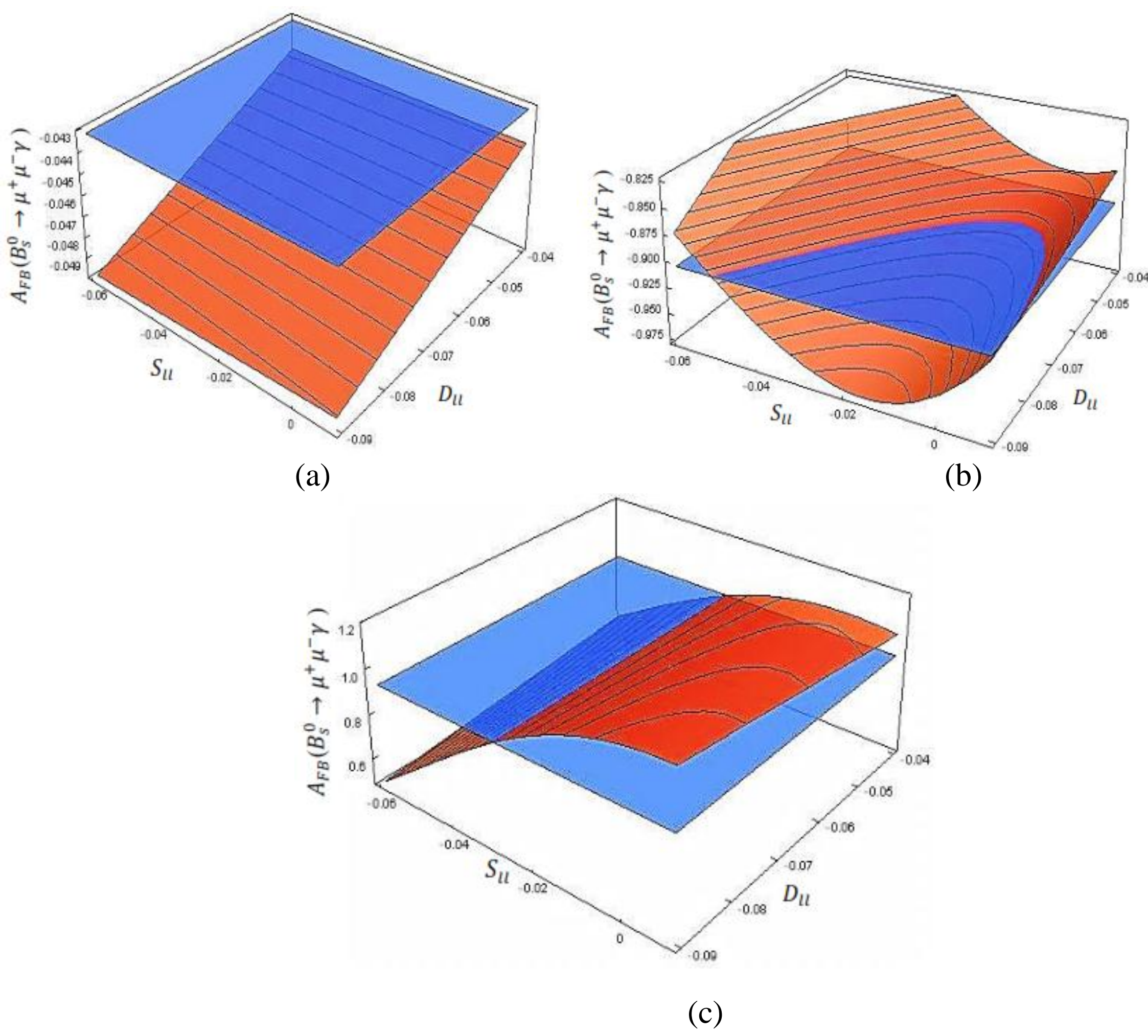

Fig. 4: Represents the dependence of $A_{FB}(B_s^0 \to \mu^+\mu^-\gamma\ )$ on $S_{ll}$ and $D_{ll}$ at (a) $s = 0.2\ GeV^2$ (b) $s = 2\ GeV^2$ (c) $s = 20\ GeV^2$, with $|B_{sb}| = 1.31 \times 10^{-3}$ and $\varphi_{sb} = -79°$ ($S_1$). The blue plane correspond to the SM results.

In Fig. 4 (a), the $A_{FB}(B_s^0 \to \mu^+\mu^-\gamma\ )$ increases gradually with $S_{ll}$ and $D_{ll}$ in low $s = 0.2\ GeV^2$ region with the coupling parameters within the range of $S_1$ but the $A_{FB}$ is less than the SM result. The sketch is quite opposite in Fig. 5 (a) and (b), here the $A_{FB}$ is considerably enhanced from the SM depending on $Z'$ coupling parameter within scenario $S_2$ in low $s = 0.2\ \text{GeV}^2$ region. In Fig. 4 (b), $A_{FB}$ slope initially drops and then enhances noticeably from the SM result at $s = 2\ \text{GeV}^2$. The situation is a bit different in Fig. 4 (c) and 5 (c) where $A_{FB}$ in high $s$ region slowly increases and crosses the SM value with increasing values of $Z'$coupling parameters in scenario $S_1$ and $S_2$ respectively. Fig. 6 and 7 show the $A_{FB}$ for $B_s^0 \to \tau^+\tau^-\gamma$ decay which is enhanced from the SM results in both the $Z'$scenario $S_1$ and $S_2$ respectively. The nature of the slopes of the $A_{FB}$ for $B_s^0 \to \tau^+\tau^-\gamma$ decay in Fig. 6 (a), (b) and 7 (a), (b) are almost similar, but they differ from slopes of the $A_{FB}$ for $B_s^0 \to \mu^+\mu^-\gamma$ decay shown in Fig. 4, Fig. 5 and Fig. 6. This may indicate towards the lepton non-universality in these decay modes. Furthermore, Sirvanh [75] studied $B_s^0 \to \ell^+\ell^-\gamma$ decays in a family non-universal $Z'$ model in different $q^2$ regions and found that the measurable quantities (e.g. differential branching ratio, double lepton polarizations and forward-backward asymmetries) are quite sensitive to the $Z'$ contributions and give indication for the existence of new physics. Recently, Abbas *et al.* [128] have studied $B_s^0 \to \ell^+\ell^-\gamma$ $(\ell = \mu, e)$ decays in the presence of additional NP vector and axial vector operators along with the charm loop effects in the form of corrections to the Wilson coefficients. From their study, they conclude that the present $b \to s\mu^+\mu^-$ data does not allow large deviation in any of the observables of $B_s^0 \to \ell^+\ell^-\gamma$ decays from their SM predictions. Though their result does not show large deviation from SM values but we can say the distinction between the results indicate that there is enough scope to excite NP effects.

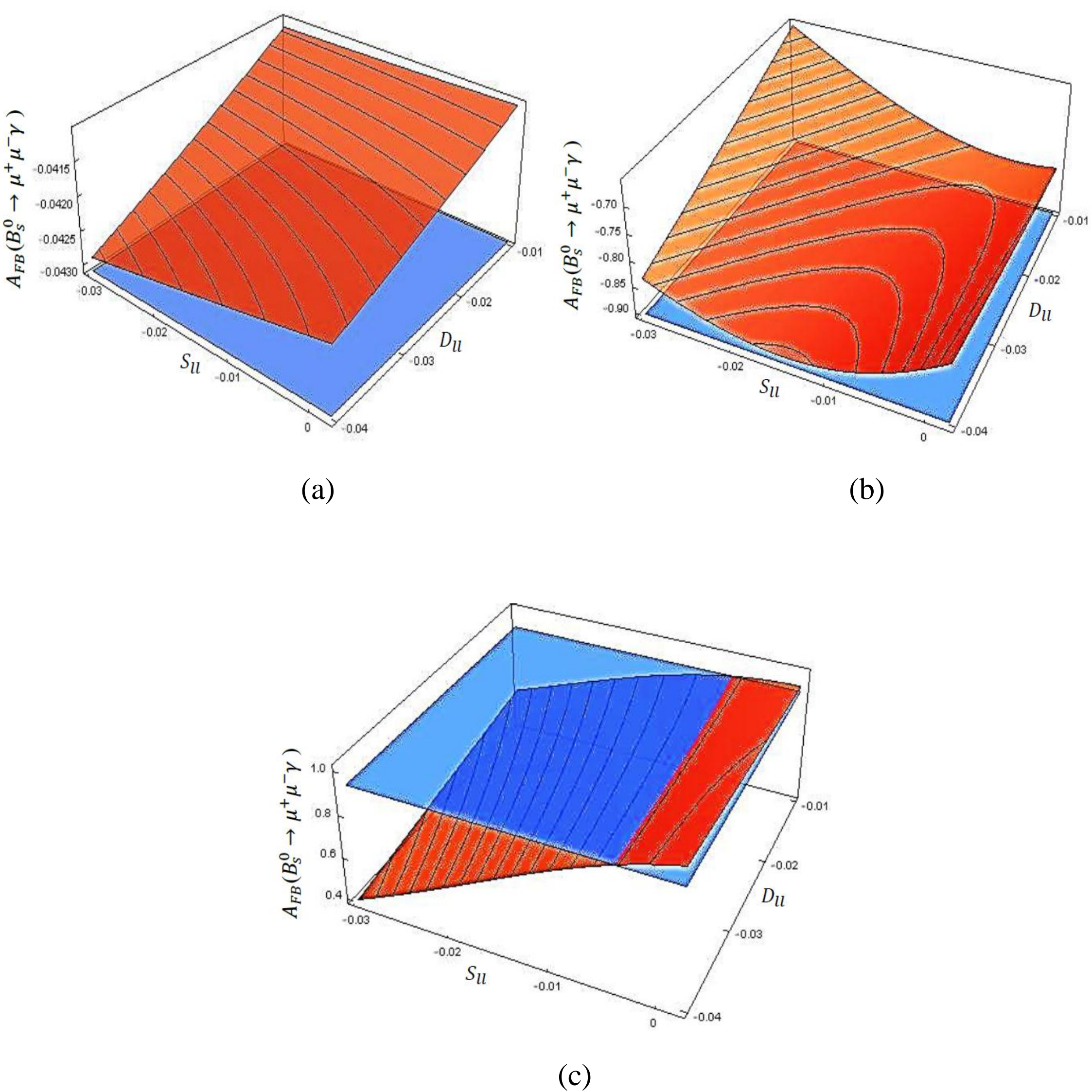

Fig. 5: Represents the dependence of $A_{FB}(B_s^0 \to \mu^+\mu^-\gamma)$ on $S_{ll}$ and $D_{ll}$ at : (a) $s = 0.\ 2GeV^2$ (b) $s = 2\ GeV^2$ (c) $s = 20\ GeV^2$, with $|B_{sb}| = 2.35 \times 10^{-3}$ and $\varphi_{sb} = -86° \ (S_2)$. The blue plane corresponds to the SM results.

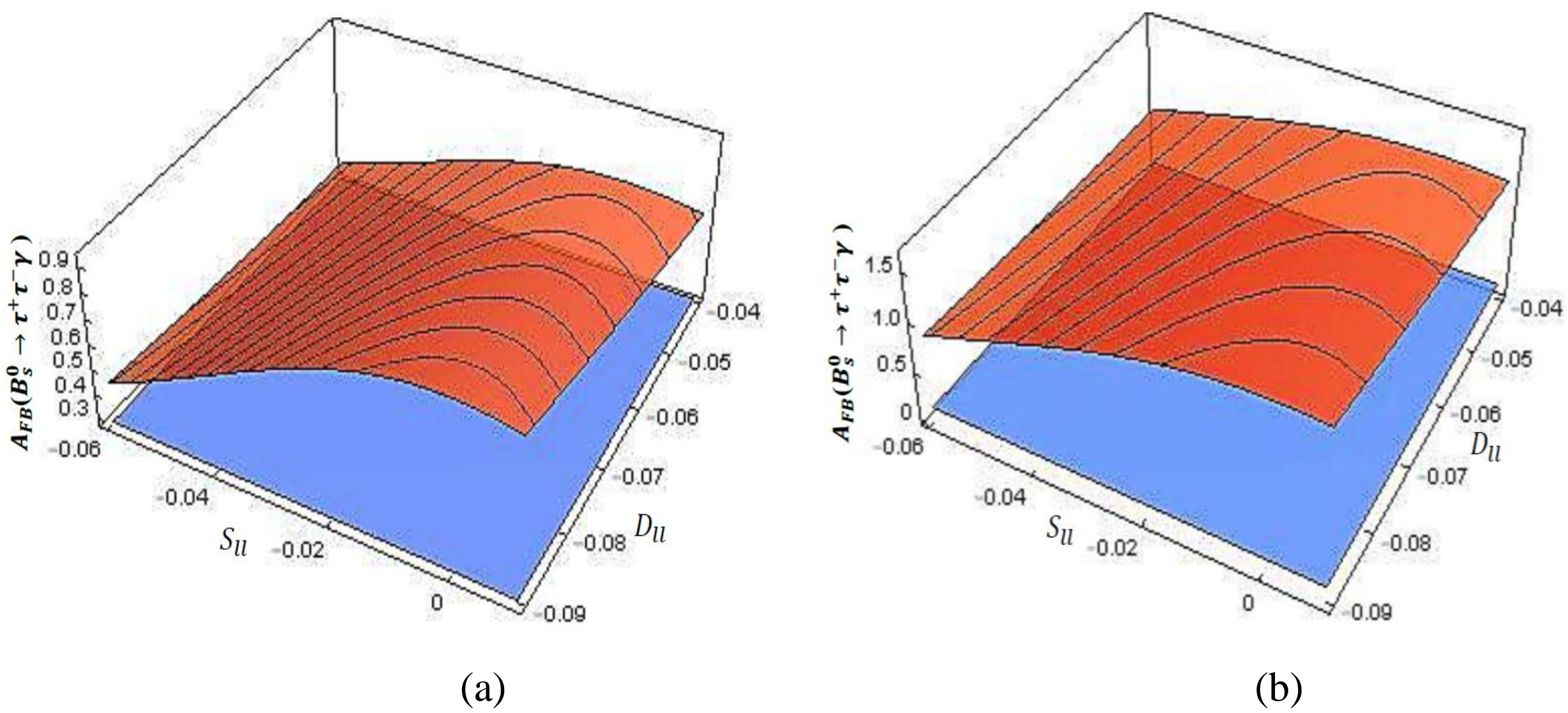

(a) (b)

Fig. 6: Represents the dependence of $A_{FB}(B_s^0 \to \tau^+\tau^-\gamma)$ on $S_{ll}$ and $D_{ll}$ at : (a) $s = 13GeV^2$ (b) $s = 25\ GeV^2$ with $|B_{sb}| = 1.31 \times 10^{-3}$ and $\varphi_{sb} = -79°\ (S_1)$. The blue plane corresponds to the SM results.

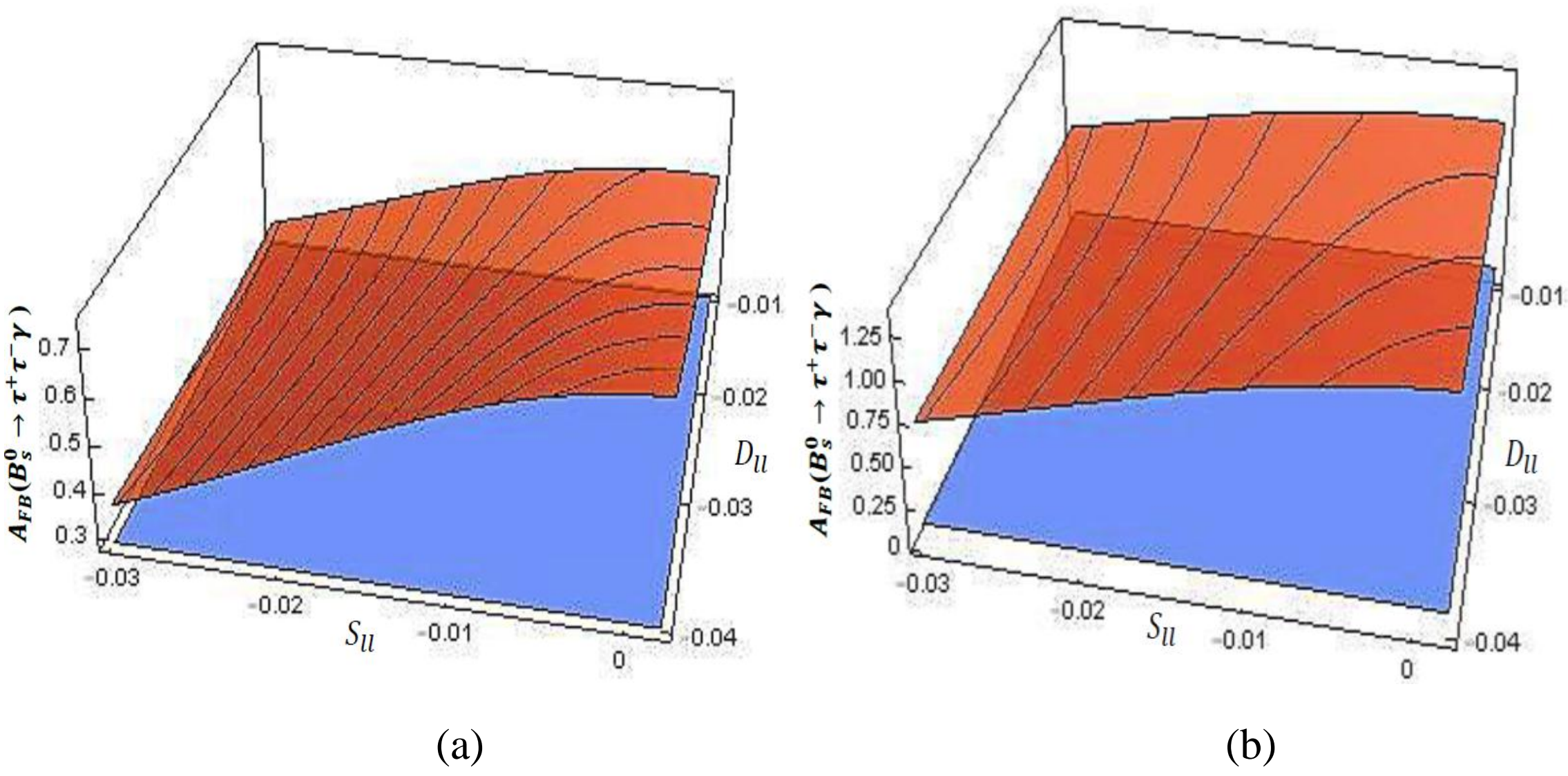

(a) (b)

Fig. **7**: Represents the dependence of $A_{FB}(B_s^0 \to \tau^+\tau^-\gamma)$ on $S_{ll}$ and $D_{ll}$ at : (a) $s = 13\text{GeV}^2$ (b) $s = 25\ \text{GeV}^2$ with $|B_{sb}| = 2.35 \times 10^{-3}$ and $\varphi_{sb} = -86°\ (S_2)$. The blue plane corresponds to the SM results.

## 5. Conclusion

In this paper, we have studied the effect of non-universal $Z'$ boson on the rare radiative leptonic decays $B_s^0 \to \ell^+\ell^-\gamma(\ell = \mu, \tau)$. This non-universal $Z'$ model allows FCNC transitions at tree level which gives a boost for the physical observables compare to their SM values. The estimated branching ratios for $B_s^0 \to \ell^+\ell^-\gamma(\ell = \mu, \tau)$ decay processes in $Z'$ model are enhanced by one order with respect to their SM predictions. The branching ratios depend on the coupling of $Z'$ boson with that of quarks and leptons. The Fig.3 depicts the same in both scenarios $S_1$ and $S_2$ with different $\varphi_{sb}$ values. The forward-backward asymmetry plays a vital role to discriminate NP effects. Our plots show that the forward-backward asymmetry for $B_s^0 \to \ell^+\ell^-\gamma$ ($\ell = \mu, \tau$) decay processes in $Z'$model depend on the $Z'$coupling parameters severely and it also clearly show substantial difference between NP contributions and SM results. These discrepancies between the prediction of the SM and the non-universal $Z'$ model may be an indication for the existence of the $Z'$ boson. In the present work, we have studied the possible NP scenario in $B_s^0 \to \ell^+\ell^-\gamma$ decays in the light of non-universal $Z'$ model but the precise measurement of observables like branching ratios and forward-backward asymmetry for $B_s^0 \to \ell^+\ell^-\gamma$ decays are very much needed to test the allowed NP models. Detection of these decays in present/future colliders with the detailed measurements and analysis of the physical observables would clear all the conjecture about the NP predictions.

**Acknowledgement**

D. Banerjee (IF140258) acknowledges the Department of Science and Technology (DST), Government of India for providing INSPIRE Fellowship. P. Nayek and S. Sahoo would like to thank SERB, DST, Government of India for financial support through the Sanction order EMR/2015/000817.